\documentclass[fleqn,10pt]{wlscirep}

\title{Theory of Multifarious Quantum Phases and Large Anomalous Hall effect in Pyrochlore Iridate Thin Films}

\author[1,*]{Kyusung Hwang}
\author[1,2,3,+]{Yong Baek Kim}

\affil[1]{Department of Physics and Centre for Quantum Materials, University of Toronto, Toronto, Ontario M5S 1A7, Canada}
\affil[2]{Canadian Institute for Advanced Research/Quantum Materials Program, Toronto, Ontario M5G 1Z8, Canada}
\affil[3]{School of Physics, Korea Institute for Advanced Study, Seoul 130-722, Korea}

\affil[*]{khwang@physics.utoronto.ca}
\affil[+]{ybkim@physics.utoronto.ca}

\begin{abstract}

We theoretically investigate emergent quantum phases in the thin film geometries of the pyrochore iridates,
where a number of exotic quantum ground states are proposed to occur in bulk materials as a result of the 
interplay between electron correlation and strong spin-orbit coupling. The fate of these bulk phases as well
as novel quantum states that may arise only in the thin film platforms, are studied via a theoretical model 
that allows layer-dependent magnetic structures. It is found that the magnetic order develop in inhomogeneous fashions
in the thin film geometries. This leads to a variety of magnetic metal phases with
modulated magnetic ordering patterns across different layers. Both the bulk and boundary electronic states
in these phases conspire to promote unusual electronic properties. In particular, such phases are akin to 
the Weyl semimetal phase in the bulk system and they would exhibit an unusually large anomalous Hall effect. 

\end{abstract}
\begin{document}

\flushbottom
\maketitle

\thispagestyle{empty}

\section*{Introduction}

As a new venue for emergent quantum phases, correlated electron systems with strong spin-orbit coupling have provided exciting playgrounds for quantum materials physics \cite{witczak2014,rau2015,schaffer2015}. Pyrochlore iridates, R$_2$Ir$_2$O$_7$ (R is a rare earth element or Y), are prominent examples of such systems and have been extensively studied in the last few years \cite{wan2011,witczak2012,hu2012,chen2015,machida2010,go2012,moon2013,pesin2010,yanagishima2001,matsuhira2007,taira2001,fukazawa2002,matsuhira2011,hasegawa2010,zhao2011,ishikawa2012,tomiyasu2012,tafti2012,sakata_2011,shapiro2012,disseler2012,disseler2012_2nd,sagayama2013,guo2013,tian2015,ueda2015,kondo2015,fujita2015,fujita2016,fujita2016_2nd,yang2010,witczak2013,yang2011,yang2014,yamaji2014,hu2015,ishizuka2012,Udagawa2012}. 
Various novel phases have been proposed to occur as the results of the interplay between electron correlation and strong spin-orbit coupling,
which include Weyl semimetal \cite{wan2011,witczak2012}, topological insulator \cite{witczak2012,hu2012}, Chern insulator \cite{hu2012,chen2015}, chiral spin liquid \cite{machida2010}, axion insulator \cite{wan2011,go2012}, non-fermi-liquid \cite{moon2013}, and topological Mott insulator \cite{pesin2010}. A great deal of experimental effort has been put forward to realize such phases in a number of pyrochlore iridate systems \cite{yanagishima2001,matsuhira2007,taira2001,fukazawa2002,matsuhira2011,hasegawa2010,zhao2011,ishikawa2012,tomiyasu2012,tafti2012,sakata_2011,shapiro2012,disseler2012,disseler2012_2nd,sagayama2013,guo2013,tian2015,ueda2015,kondo2015,fujita2015,fujita2016,fujita2016_2nd}.

Previous theoretical studies have shown that the emergence of many novel phases can be understood by starting from a quadratic band touching (QBT) semimetal in the paramagnetic state, where two doubly-degenerate bands touch quadratically at the zone center \cite{yang2010}.
Upon increasing the interaction or the Hubbard $U$, the system develops the all-in/all-out (AIAO) magnetic order, which breaks the time reversal symmetry, but not the cubic symmetry of the lattice \cite{wan2011,witczak2012,go2012,witczak2013}. This results in the splitting of the previously doubly-degenerate bands. Then the crossing of two non-degenerate bands leads to the Weyl semimetal (WSM) phase, where a pair of Weyl fermions and their cubic symmetry equivalents appear \cite{wan2011,witczak2012,go2012,witczak2013}. Further increasing the interaction makes a pair of Weyl fermions to move into high symmetric points in the Brillouin zone and when they meet, a band gap opens up at the high symmetry points, leading to a fully gapped magnetic insulator. Most of the pyrochlore iridates show a finite temperature (semi)metal-insulator transition and develop a magnetic order in the low temperature insulating phase while Pr$_2$Ir$_2$O$_7$ is the only member of the family, that remains paramagnetic down to a few Kelvin \cite{yanagishima2001,matsuhira2007,matsuhira2011}.
Various experiments on R=Eu, Y, Nd compounds show evidences for the AIAO magnetic order \cite{zhao2011,tomiyasu2012,sagayama2013}.
Remarkably, a recent angle-resolved photo-emission spectroscopy (ARPES) measurement shows that the paramagnetic state of Pr$_2$Ir$_2$O$_7$ has the 
QBT at the zone center, confirming the basic theoretical picture \cite{kondo2015}. On the other hand, a clear evidence for the Weyl semimetal phase in this class of materials is currently lacking. From the theoretical point of view, the region in the bulk phase diagram where the Weyl semimetal phase exists is limited to a narrow window of intermediate strength of the Hubbard $U$ \cite{witczak2012,go2012,witczak2013}, and it is perhaps difficult to identify such a phase in the bulk materials. 

\begin{figure}
\centering
\includegraphics[width=\textwidth]{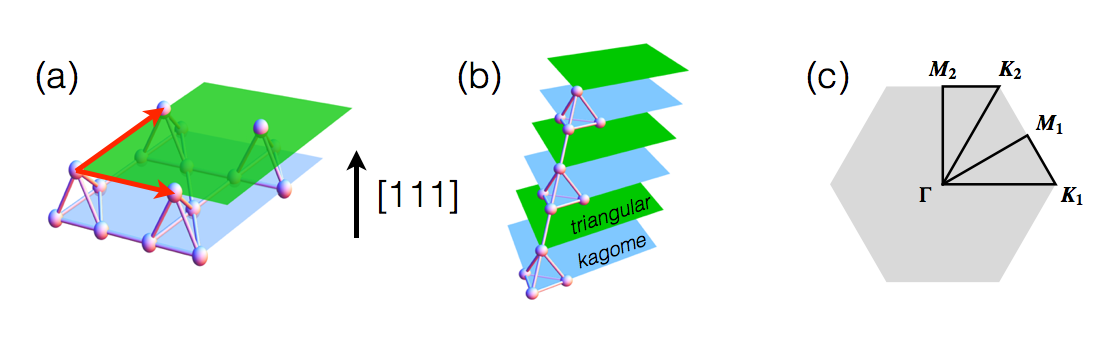}
\caption{Lattice structure of the Ir$^{4+}$ ions in the [111] pyrochlore thin films.
(a) Bilayer structure formed with the kagome (light blue) and triangular (green) layers of the Ir sites (balls). (b) Stacking of the bilayers. For simplicity, the figure only shows three bilayers and the sites in a corresponding unit cell. (c) The Brillouin zone of the pyrochlore film structures.
Red arrows in (a) represent the lattice vectors in the film structures.
\label{fig:filmstructure}}
\end{figure}

In this work, we theoretically explore the emergence of novel magnetic metallic phases akin to the Weyl semimetal \cite{burkov2011,wan2011,witczak2012}, in thin film geometries of the pyrochlore iridates.
The primary motivations for this study are twofold. First, one of the prominent signatures of the Weyl fermions in the bulk system is the presence of the Fermi arc at the surface \cite{wan2011,witczak2012}, which is rather difficult to study in the three-dimensional cubic crystal. In this thin film geometries, both of the surface and bulk states may be more easily probed at the same time. Secondly, another important signature of the Weyl fermions is the large anomalous Hall effect. When a pair of Weyl fermions are present, the Hall conductivity is of the order of $e^2/h$ or proportional to $e^2/h$ times the distance between two Weyl points in the Brillouin zone \cite{yang2011}. In the cubic crystal, there also exist other Weyl points related by the cubic symmetry, which leads to the cancellation of the Hall conductivity and zero anomalous Hall effect. Hence the broken cubic symmetry of the thin film geometries may resurrect the large anomalous Hall effect by nullifying the cancellation \cite{yang2011}. 

We consider various pyrochlore film geometries with the boundary surfaces along the [111] direction and a realistic film thickness.
Employing the effective tight-binding model \cite{witczak2013} for the $j_{\textup{eff}}=1/2$ states from Ir$^{4+}$ ions on the thin film geometry,
we explore topological and magnetic phases fostered by the interplay between electron correlation, boundary surfaces, and lowered symmetries.
We choose a parameter regime where the corresponding bulk system exhibits (i) the semimetal with the QBT at the zone center in the paramagnetic state (ii) 
the Weyl semimetal at intermediate strength of the interaction, and (iii) a gapped magnetic insulator with the AIAO order in the large interaction limit. 
Both the layer-dependance of the magnetic order parameter and the corresponding electronic structure are investigated by performing the self-consistent
unrestricted Hartree-Fock mean-field theory computations, where the magnetic moments in each layer are allowed to point in any direction.

An important discovery in this work is the appearance of the modulated magnetic order in the thin film geometries, where the amplitude and canting angles of the magnetic moments change from one layer to the other. The magnetic order can be regarded as a canted version of the AIAO order with varied canting angles depending on the layers.
As the interaction strength increases, the magnetic order develops at the boundary layers first and then spreads to the central parts of the film. In general, the magnetic orders at the boundary layers are stronger than those near the central layers in the weak to intermediate correlation regime. This means the effective electron correlation is bigger at the boundary compared to the ones near the central layers. For example, there is a regime where only the boundary layers are magnetically ordered and the central layers remain paramagnetic. This is an interesting situation where only thin film system may realize various different parts of the bulk phase diagram in different locations of the layers.

Because of the modulated magnetic order, the resulting electronic structure reflects intricate combinations of the bulk (near the central layers) and boundary states. One of the most interesting phases in the thin film systems is the magnetic metal with the modulated magnetic order in the intermediate correlation regime, where the splitting of the QBT occurs via the inhomogeneous effective fields from different layers with varied strength of the magnetic order. Here a plethora of crossings of non-degenerate bands lead to a complex structure of the Berry flux with many ``monopoles" and ``anti-monopoles" in momentum space. These crossings represent the thin film version of the Weyl fermions and lead to a large anomalous Hall effect as the cubic symmetry is explicitly broken in the thin film geometries. 
We also demonstrate the appearance of a variety of inhomogeneous magnetic metal phases depending on the choice of the boundary layers.
This raises the hope that one could control the nature of the bulk and surface states by making appropriate choices of the boundary lattices.

It should be pointed out that the appearance of the modulated or inhomogeneous magnetic order is crucial for the Berry phase structure with many ``monopoles"
and ``anti-monopoles", and the resulting large anomalous Hall effect. In a previous study\cite{yang2014} of the thin film geometries,
the uniform magnetic order was assumed and the largest anomalous Hall effect was found to occur in the so-called ``hidden topological phase". Such a phase
exists only for specific choices of boundary layers. As explained in detail in the main text, our study shows that the underlying physics is very different 
when the inhomogeneous magnetic order arises and the largest possible anomalous Hall effect can occur irrespective of the presence of such phases or
different choices of boundary layers.

\section*{Results}

\begin{figure}
\centering
\includegraphics[width=\textwidth]{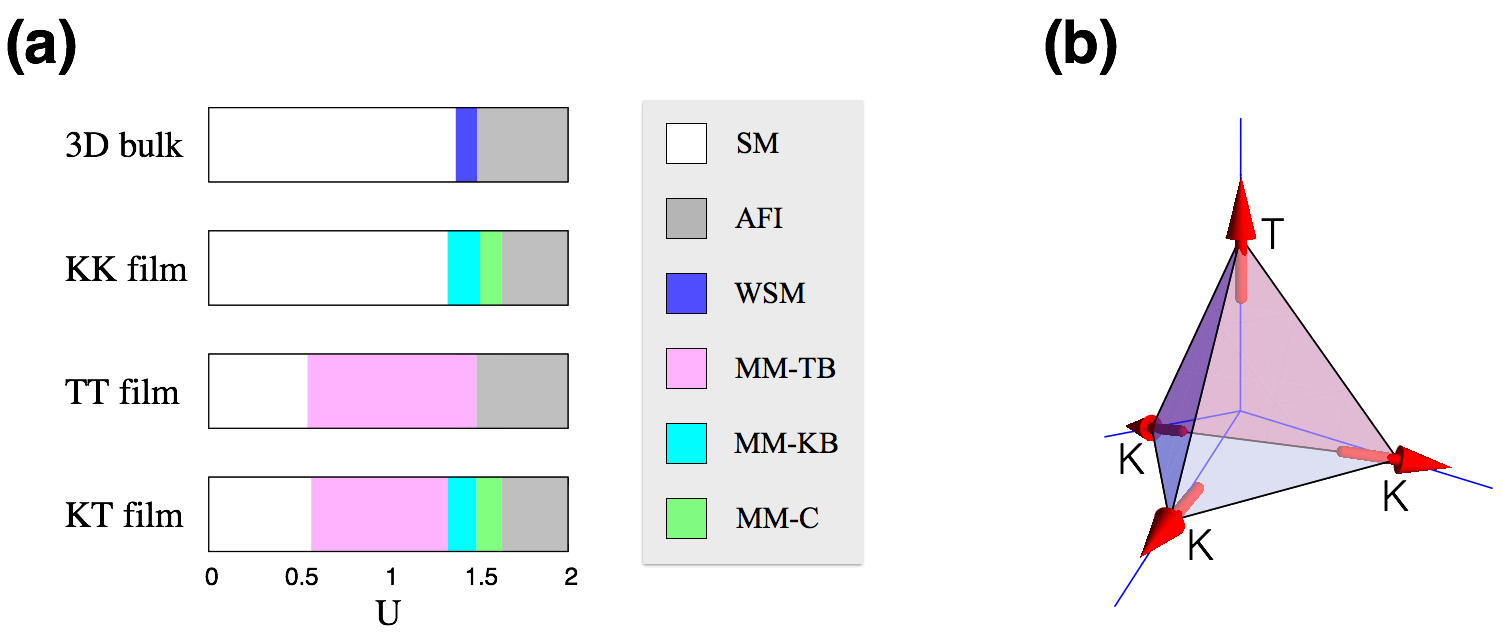}
\caption{Phases of the theoretical model in Eq. (\ref{eq:tight-binding}).
(a) Mean-field phase diagrams for the 3D bulk system and [111] thin films with twenty bilayers. The diagrams include the semimetal (SM; white), antiferromagnetic insulator (AFI; gray), Weyl semimetal (WSM; blue), and a variety of magnetic metal (MM; pink, cyan, green) phases. For the classification of the MM phases, see the main text.
(b) AIAO magnetic ordering pattern in the AFI and MM phases.
In the tetrahedron, the bottom three and top single sites belong to the kagome and triangular layers as labeled by K and T.
The magnetic moments $\{\langle {\bf J}_i \rangle\}$ (red arrows) point along the local axes (blue lines) with slight canting of the moments at the kagome sites.
\label{fig:phasediagrams}}
\end{figure}

\subsection*{A. Theoretical Model}

We consider a generic model for the thin films of the pyrochlore iridates grown along the [111] direction, which has the structure of alternating kagome and triangular layers of Ir$^{4+}$ ions. Such structures can be regarded as the stackings of the kagome-triangular-bilayers [Fig. \ref{fig:filmstructure} (a) and (b)]. Only the lattice sites in a unit cell are shown in Fig. \ref{fig:filmstructure} (b) for simplicity while the kagome and triangular layers are schematically represented by light blue and green sheets, respectively. In the thin film structures, there are three distinct choices for the top and bottom boundary surfaces: (i) kagome and kagome (KK) layers, (ii) triangular and triangular (TT) layers, and (iii) kagome and triangular (KT) layers. Regardless of the boundary choice, the [111] film geometry has the $C_3$ rotation symmetry along the [111] axis. For the KK and TT boundary conditions, the system additionally has the inversion symmetry.

To investigate electronic and magnetic structures allowed in the [111] pyrochlore films, we employ the following tight binding model (introduced in Ref. \citeonline{witczak2013}) based on the $j_{\textup{eff}}=1/2$ Kramers doublet\cite{kim2008,kim2009} of the Ir$^{4+}$ itinerant electrons.
\begin{equation}
H
=
\sum_{(i,j) \in \textup{NN}} c_i^{\dagger} (t_1 \mathbb{I}_2 + i t_2 {\bf d}_{ij} \cdot \boldsymbol{\sigma} ) c_j
+
\sum_{(i,j) \in \textup{NNN}} c_i^{\dagger} (t'_1 \mathbb{I}_2 + i [ t'_2 {\bf R}_{ij} + t'_3 {\bf D}_{ij} ] \cdot \boldsymbol{\sigma} ) c_j 
+
\frac{U}{2} \sum_i n_{i} (n_{i}-1),
\label{eq:tight-binding}
\end{equation}
where $c_i^{\dagger}=(c_{i\uparrow}^{\dagger},c_{i\downarrow}^{\dagger})$ and $c_j=(c_{j\uparrow},c_{j\downarrow})^T$ are creation and annihilation operators for the $j_{\textup{eff}}=1/2$ states with the pseudospin $\uparrow$, $\downarrow$. Here the $j_{\textup{eff}}=1/2$ band is half-filled.
The first two terms in the model Hamiltonian describe the symmetry-allowed electron hopping processes for the nearest-neighbours (NN) and next nearest-neighbours (NNN) with the spin-independent ($t_1$, $t'_1$), and spin-dependent hopping processes ($t_2$, $t'_2$, $t'_3$). The two-by-two matrices, $\mathbb{I}_2$ and $\boldsymbol{\sigma}=(\sigma^x,\sigma^y,\sigma^z)$ are the identity and Pauli matrices. Readers interested in more details of the tight binding model including the 3D pseudo-vectors, ${\bf d}_{ij}$, ${\bf R}_{ij}$, and ${\bf D}_{ij}$, are referred to Ref. \citeonline{witczak2013}. The last term, expressed in terms of the electron density $n_i=c_i^{\dagger} \mathbb{I}_2 c_i$, represents the Hubbard on-site repulsion ($U$) to incorporate electron correlation. We choose the parameter regime: $t_{oxy}=1$, $t_{\sigma}=-0.8$, $t_{\pi}=-2t_{\sigma}/3$, and $t'_{\sigma}/t_{\sigma}=t'_{\pi}/t_{\pi}=0.08$ in the Slater-Koster parametrization for the hopping amplitudes of Ref. \citeonline{witczak2013}. In the 3D bulk system with the same parameters, a mean-field theory leads to the SM with the QBT, WSM, and AFI with the AIAO order upon controlling $U$ [see the 3D bulk phase diagram in Fig. \ref{fig:phasediagrams} (a)].

\begin{figure}
\centering
\includegraphics[trim= 0mm 40mm 0mm 40mm, clip, width=\linewidth]{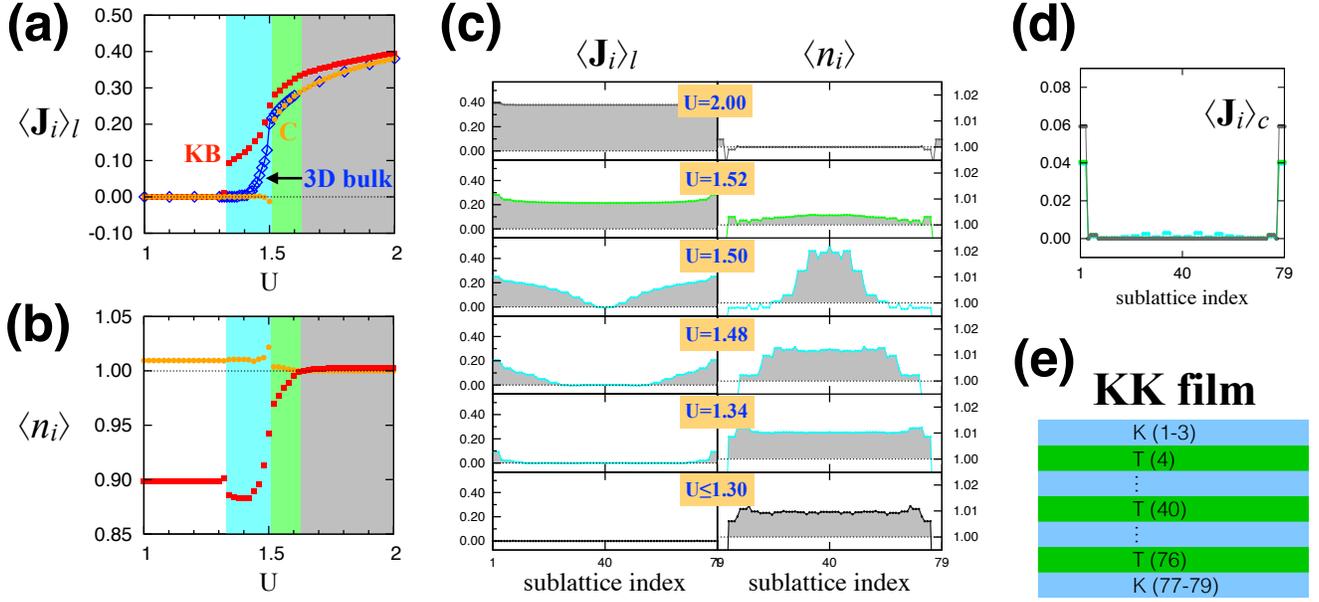}
\caption{Magnetic structure of the KK film.
(a) The local axis component $\langle {\bf J}_i \rangle_l$ of the magnetic order parameter. 
The plot shows $\langle {\bf J}_i \rangle_l$ as a function of $U$ at the kagome boundary (KB; red) and central layer (C; orange) in the film.
For comparison, the magnetic order parameter for the 3D bulk system (3D bulk; blue) is plotted together.
(b) The electron density $\langle n_i \rangle$ with the same color scheme in (a).
(c) Spatial profiles of $\langle {\bf J}_i \rangle_l$ and $\langle n_i \rangle$ over the layers (sublattices) in the KK film.
(d) Spatial profiles of the canting component $\langle {\bf J}_i \rangle_c$ of the magnetic order parameter for $U$=1.50 (cyan), 1.52 (green), 2.00 (gray).
(e) Sublattice index for the KK film with twenty bilayers. There are three sublattices on each kagome layer while only one on a triangular layer [see Fig. \ref{fig:filmstructure} (b)].
\label{fig:kkfilm1}}
\end{figure}

In this parameter regime, we construct a self-consistent mean-field theory for the [111] pyrochlore thin film with 20 bilayers for the three different boundary conditions: KK, TT, and KT. In this mean-field theory, the Hubbard interaction term is decoupled via $\frac{1}{2} \sum_i n_{i} (n_{i}-1) \rightarrow -\frac{4}{3} \langle {\bf J}_i \rangle \cdot {\bf J}_i + \frac{2}{3} \langle {\bf J}_i \rangle^2$ with ${\bf J}_i = \frac{1}{2} c_i^{\dagger} \boldsymbol{\sigma} c_i$, and we allow layer-dependent configurations $\langle {\bf J}_i \rangle$ of the magnetic order. 
We solve the resulting mean-field Hamiltonian self-consistently for each boundary condition.
The mean-field phase diagrams obtained for different boundary conditions are plotted in Fig. \ref{fig:phasediagrams} (a). We find three different classes of phases in the film systems: nonmagnetic semimetal (SM), antiferromagnetic insulator (AFI), and a variety of magnetic metal (MM) phases.
In the subsequent sections, we explain the nature of these phases and describe their properties by focusing first on the results for the KK film shown in Figs. \ref{fig:kkfilm1} and \ref{fig:kkfilm2}. The cases for other boundary conditions are then discussed in comparison with the KK film.

\subsection*{B. Magnetic Structure}

We first describe the nature of different phases by investigating the magnetic properties.
As in the 3D bulk system, the SM and AFI phases [white and grey in Fig. \ref{fig:phasediagrams} (a)] appear in the weak and strong correlation regimes, respectively, commonly for all three boundary conditions. A variety of magnetic metallic (MM) phases appear in the intermediate correlation regime.
Details of the magnetic structures in these phases are explained below. 

\subsubsection*{Semimetal}
First, the nonmagnetic SM phase in the weak correlation regime is characterized by zero magnetic order parameter, $\langle {\bf J}_i \rangle=0$.
As will be discussed later, the electronic band structure of the SM can be regarded as the film version (finite size along the [111] direction) of 
the quadratic band touching at the $\Gamma$ point as seen in the corresponding 3D bulk phase.

\subsubsection*{Antiferromagnetic insulator}
The AFI phase in the strong correlation regime overall has the AIAO magnetic ordering pattern with slight modifications.
The modifications occur in two different ways: (i) canting of the moments from the ideal AIAO ordering pattern, and 
(ii) layer-dependent modulation in the magnitude of the magnetic order parameter $\langle {\bf J}_i \rangle$.
The canting of the moments off the local axis arises on the kagome layers while the moments on the triangular layers are strictly along 
the local axis directions [Fig. \ref{fig:phasediagrams} (b)].
To characterize the canting of the moments, we decompose the magnetic order parameter $\langle {\bf J}_i \rangle$ into two parts:  the component along the local axis ($\langle {\bf J}_i \rangle_l$) and the perpendicular component representing the canting ($\langle {\bf J}_i \rangle_c$).
The left column of Fig. \ref{fig:kkfilm1} (c) and Fig. \ref{fig:kkfilm1} (d) show these components of the moments as functions of the 
sublattice index and $U$ for the KK film. 
The sublattices are numbered sequently from the top to the bottom layers as shown in Fig. \ref{fig:kkfilm1} (e).
Note that there are three sublattices on each kagome layer while only one on a triangular layer [Fig. \ref{fig:filmstructure} (b)].
By comparing Figs. \ref{fig:kkfilm1} (c) and (d), we find that the canting component $\langle {\bf J}_i \rangle_c$ is substantially small compared to the major component $\langle {\bf J}_i \rangle_l$ along the local axis.
In the case of $U=2$ (gray), we clearly see the dominance of $\langle {\bf J}_i \rangle_l~(\simeq 0.40)$ over $\langle {\bf J}_i \rangle_c~(\lesssim 0.06)$.
As shown in the left column of Fig. \ref{fig:kkfilm1} (c), the size of the magnetic moment is modulated in different sublattices/layers.
In general, magnetic order is stronger near the boundary layers rather than the central layers.
This pattern is more evident in the MM phases [$U=1.34,~1.48,~1.50$ in Fig. \ref{fig:kkfilm1} (c) left column] as explained further below.
In spite of the canting of the moments and modulated magnetic order, it is found that the magnetic order in the film systems preserves 
the $C_3$ rotation symmetry regardless of the boundary condition.
In the KK and TT films, the inversion symmetry is also preserved in the presence of the magnetic order [Figs. \ref{fig:kkfilm1} (c) and (d)].

\subsubsection*{Magnetic metals}
In the intermediate correlation regime, three distinct MM phases emerge as the electron correlation ($U$) increases in the film systems
with various boundary conditions. The canting of the moments and the modulation of the magnetic order are more pronounced in this regime
compared to the AFI phase in the strong correlation regime.
Based on their magnetic structures at the boundary and central layers, we classify them into three different categories: (i) MM-TB, (ii) MM-KB, and (iii) MM-C.
First, in the MM-TB and MM-KB phases, the magnetic order is present mainly around the boundary layers while the central layers are mostly non-magnetic. 
Here, the suffixes TB/KB mean that the triangular/kagome boundary layer has the magnetic order, respectively.
In the MM-C phase, the magnetic order prevails in the central layers as well as near the boundaries.
In general, the phases MM-TB, MM-KB, and MM-C appear sequently as the Hubbard $U$ increases [see Fig. \ref{fig:phasediagrams} (a)].
More specifically, in the KK film, the MM-KB phase is found in the region $1.33 < U < 1.51$ and the MM-C phase in $1.51 < U < 1.63$.
As shown in the left column of Fig. \ref{fig:kkfilm1} (c), the magnetic order is first developed from the kagome boundaries and then 
spreads towards the central layers (the cases of $U=1.34,1.48,1.50$ for the MM-KB phase) as the Hubbard $U$ increases.
At $U=1.52$, the magnetic order at the central layers suddenly jumps from nearly zero to a finite size (0.2), leading to the MM-C phase.
Similar trends can be seen in the cases of the TT and KT films. In the TT film, the transition from the MM-TB to the AFI occurs
without going through the MM-C phase (Fig. \ref{fig:ttfilm1}). That is, when the magnetic order spreads to the central layers, the bulk spectrum becomes 
immediately gapped. In the case of the KT film, the triangular boundary layers develop the magnetic order first (MM-TB), then the kagome boundary
layers (MM-KB), and finally the central layers acquire the magnetic order (MM-C) as the correlation $U$ increases (Fig. \ref{fig:ktfilm1}). 

\subsubsection*{Phase transitions}
Figure \ref{fig:kkfilm1} (a) summarizes the phases found in the KK film with the behaviors of the magnetic order 
at the kagome boundaries (KB; red) and central layers (C; orange) as functions of $U$. 
Finite jumps of the order parameters in the KB and C layers indicate the onsets of the MM-KB and MM-C phases, respectively.
In terms of the magnetic structure, the MM-C and AFI phases are essentially same. 
The only difference is the absence/presence of an energy gap in the electronic band structure.

\subsection*{C. Electronic Structure}

\begin{figure}
\centering
\includegraphics[trim= 0mm 80mm 0mm 0mm, clip, width=\linewidth]{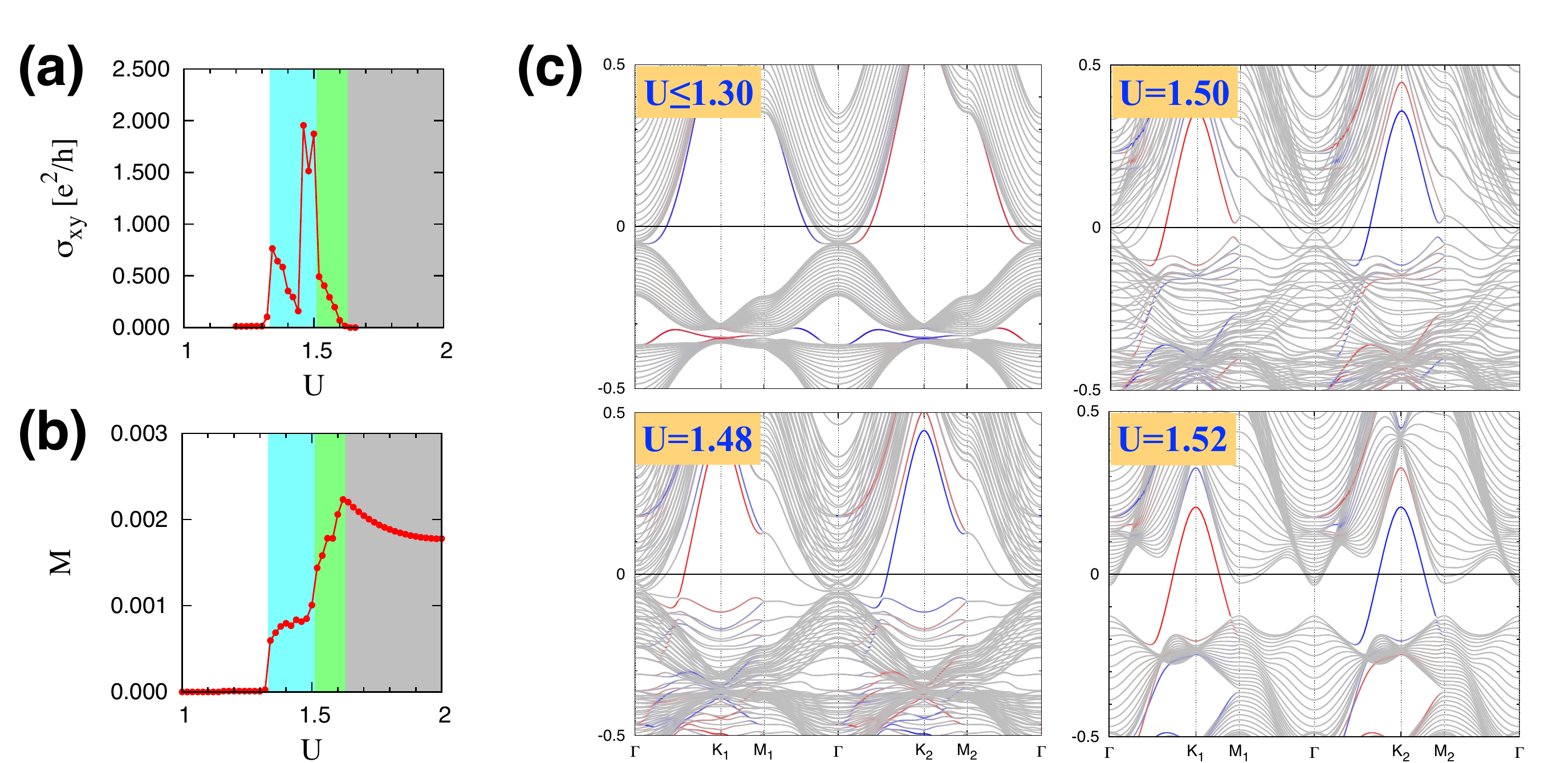}
\caption{Electronic structure of the KK film.
(a) The Hall conductivity $\sigma_{xy}$ as a function of $U$.
(b) The magnetization $M$ along the [111] direction as a function of $U$. 
(c) Electron band structures of the SM ($U\leq1.30$), MM-KB ($U=1.48,~1.50$), and MM-C ($U=1.52$) phases.
In each plot, the surface modes localized at the top and bottom kagome layers (red and blue, respectively) are distinguished from the bulk bands (gray).
\label{fig:kkfilm2}}
\end{figure}

Now we analyze the electronic structures of different phases.
Figure \ref{fig:kkfilm2} (c) shows the band structures for several cases in the KK-film system: $U\leq1.30$ (SM), $U=1.48,~1.50$ (MM-KB), and $U=1.52$ (MM-C).
In the band structure plots, we distinguish the surface modes localized at the top and bottom kagome layers (red and blue) from the bulk bands (grey).
Notice that there is no direct bulk band gap in the SM and MM-KB phases.
On the other hand, the direct band gap opens up in the MM-C phase (even though the Fermi level does not necessarily lie within the bulk band gap).
In all of the SM, MM-KB, and MM-C phases, the surface modes cross the Fermi level.
Both of the bulk and surface spectra are completely gapped in the AFI phase.

As shown in the plot for $U\leq1.30$, the SM phase has the quadratic band touching structure at the $\Gamma$ point. 
In reality, there exists a tiny energy gap between the lowest convex ($\cup$) and highest concave ($\cap$) bulk bands due to the finite thickness of the film.
The Fermi level, denoted with the zero energy, is above the touching point unlike the 3D bulk system, which happens because of the added structure of 
the surface bands. 
Both of the bulk and surface bands have twofold degeneracy at each crystal momentum due to the time reversal and inversion symmetries. 
The exactly same band structures along $\Gamma$-K$_1$-M$_1$-$\Gamma$ and $\Gamma$-K$_2$-M$_2$-$\Gamma$ reflect 
the $C_3$ rotation and inversion symmetries.

In the MM-KB phase with the broken time reversal symmetry (see the plots for  $U$=1.48 and 1.50), 
the modulated magnetic order across the layers provides a range of effective exchange field for the bulk bands
and hence the splitting of the bulk bands occur in a complex manner, resulting in a large number of 
crossings of non-degenerate bands [Fig. \ref{fig:kkfilm2} (c)].

When the magnetic order finally reaches the central layer around $U$=1.52, the system enters the MM-C phase.
Now the band structure of this phase is qualitatively different from that of the MM-KB phase [compare the cases $U=1.52$ and $1.50$ in Fig. \ref{fig:kkfilm2} (d)].
In the MM-C phase, the multitudes of band crossings disappear and the bulk band acquires a band gap while the Fermi level crosses
the protruding surface modes (see the red and blue bands crossing the Fermi level in the case of $U=1.52$).
Upon increasing $U$ further, the system enters the AFI phase, where the bulk valence bands and surface modes 
connected to them lie below the Fermi level, so the entire spectrum is gapped.

\subsection*{D. Large Anomalous Hall Effect}
One of the most interesting properties of the MM phases is the anomalous Hall effect (AHE) \cite{Jungwirth2002,onoda2002,fang2003,nagaosa2010,xiao2010,xiao2011}.
The spin-orbit coupling encoded in the tight binding model [Eq. (\ref{eq:tight-binding})] enables electrons to see the magnetically-ordered moments 
as an effective magnetic field. As a result, these phases allow a finite Hall conductivity without an external magnetic field.
For the MM phases, we study the AHE by evaluating the Hall conductivity $\sigma_{xy}$.
\begin{equation}
\sigma_{xy}=\frac{e^2}{h}\sum_{\bf k}\sum_n f_n({\bf k}) B_n^z({\bf k}) .
\end{equation}
The above expression describes the Hall conductivity in terms of the Berry curvature 
${\bf B}_n({\bf k})= \nabla_{\bf k} \times {\bf A}_n({\bf k})$, where ${\bf A}_n({\bf k})=-i\langle n{\bf k}|\nabla_{\bf k} |n{\bf k} \rangle$
is the Berry connection. Here $| n {\bf k}\rangle$ is the Bloch state and $f_n({\bf k})(=0,1)$ represents the electron occupation.

The computed Hall conductivity is plotted in Fig. \ref{fig:kkfilm2} (a) in the unit of $e^2/h$.
As expected, $\sigma_{xy}$ has finite values in the MM-KB (cyan) and MM-C (green) phases.
It is remarkable that $\sigma_{xy}$ is quite large, about two times of $e^2/h$, around the phase boundary in the MM-KB side ($1.46 \leq U \leq 1.50$).
Notice that this is the parameter regime where numerous band crossings of non-degenerate bands are generated.
Such band crossings would correspond to the monopoles/anti-monopoles for the Berry curvature \cite{onoda2002,fang2003}.
A plethora of the monopoles/anti-monopoles generated in the MM-KB phase result in the large Hall conductivity.
In Fig. \ref{fig:kkfilm2} (a), the relatively small Hall conductivity in other regions is attributed to the lack of such band crossings.

We also investigate the relationship between the Hall conductivity and the net magnetization.
We find that a finite, but small, net magnetization along the [111] direction is established as the system enters the magnetically ordered state.
Figure \ref{fig:kkfilm2} (b) shows the magnitude of the net magnetization ${\bf M}=\frac{2}{N}\sum_i \langle {\bf J}_{i} \rangle$ (where $N$ is the number of sites) as a function of the Hubbard $U$.
Comparison with the Hall conductivity [Fig. \ref{fig:kkfilm2} (a)] reveals a nontrivial and nonlinear relationship between $\sigma_{xy}$ and $M$ in the magnetic metal phases.

\begin{figure}
\centering
\includegraphics[trim= 0mm 80mm 0mm 0mm, clip, width=\linewidth]{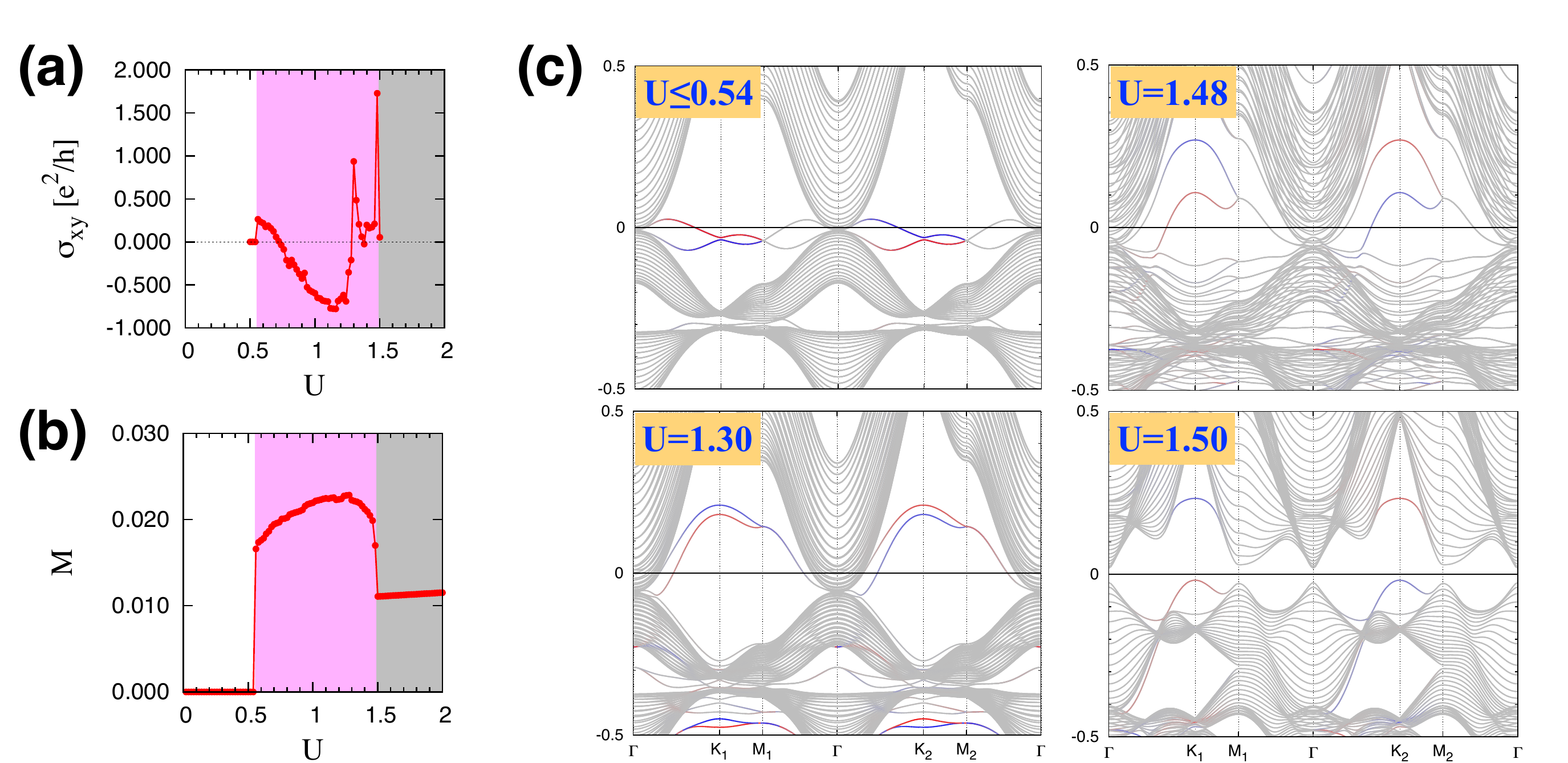}
\caption{Electronic structure of the TT film.
(a) The Hall conductivity $\sigma_{xy}$ as a function of $U$.
(b) The magnetization $M$ along the [111] direction as a function of $U$. 
(c) Electron band structures in the SM ($U\leq0.54$), MM-TB ($U=1.30,~1.48$), and AFI ($U=1.50$) phases.
In each plot, the surface modes localized at the top and bottom triangular layers (red and blue, respectively) are distinguished from the bulk bands (gray).
\label{fig:ttfilm2}}
\end{figure}

The aforementioned large Hall conductivity in the MM phases and nontrivial relationship between $\sigma_{xy}$ and $M$ are also found in the TT and KT films (see Figs. \ref{fig:ttfilm2} and \ref{fig:ktfilm2}).
On the other hand, there are several notable features that differentiate the TT and KT films from the KK film.
First, in the MM-TB phase (marked by pink) existing in the TT and KT films, the Hall conductivity changes the sign as the Hubbard $U$ increases.
Secondly, the net magnetization has a larger value in the two films compared to the KK film.
Both features originate from the magnetically-ordered triangular boundary.

\begin{figure}
\centering
\includegraphics[trim= 0mm 40mm 0mm 40mm, clip, width=\linewidth]{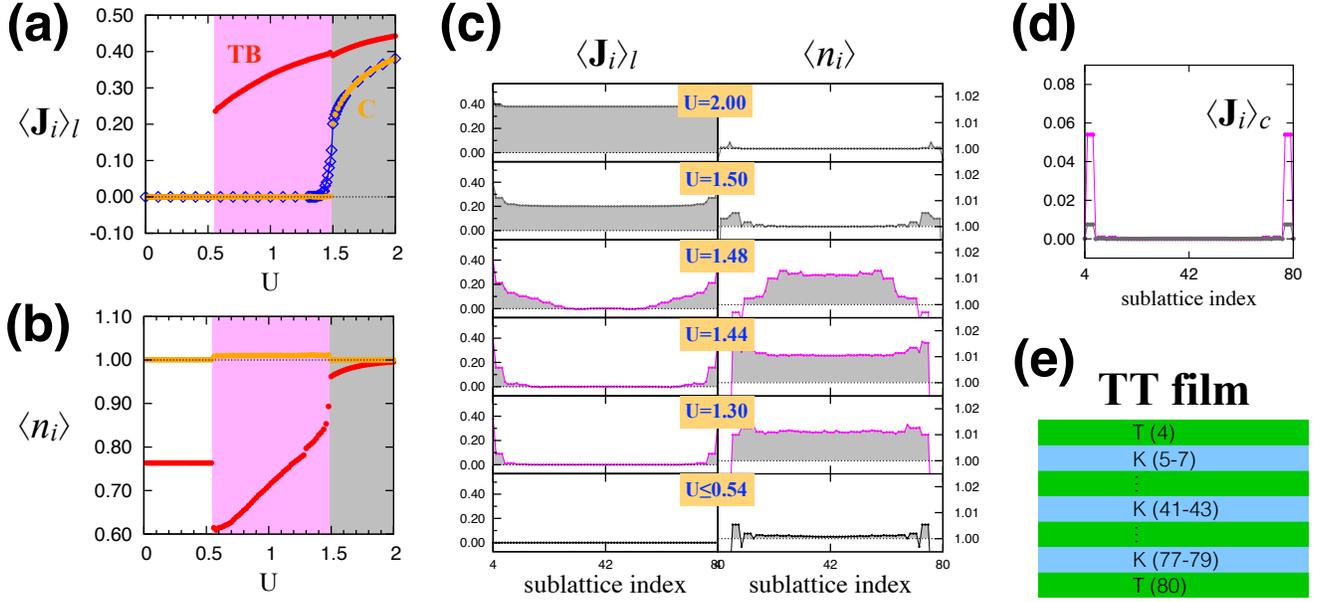}
\caption{Magnetic structure of the TT film.
(a) The local axis component $\langle {\bf J}_i \rangle_l$ of the magnetic order parameter. 
The plot shows $\langle {\bf J}_i \rangle_l$ as a function of $U$ at the triangular boundary (TB; red) and central layer (C; orange) in the film.
(b) The electron density $\langle n_i \rangle$ with the same color scheme in (a).
(c) Spatial profiles of $\langle {\bf J}_i \rangle_l$ and $\langle n_i \rangle$ over the layers (sublattices) in the TT film.
(d) Spatial profiles of the canting component $\langle {\bf J}_i \rangle_c$ of the magnetic order parameter for $U$=1.30 (pink), 2.00 (gray).
(e) Sublattice index for the TT film with twenty bilayers.
\label{fig:ttfilm1}}
\end{figure}

\section*{Discussion}
It is shown in the previous section that the AIAO magnetic order is modulated across the layers in the thin film geometries and it is generally stronger at the boundary layers compared to the central layers. At the boundary layer, the number of hopping paths is reduced due to the lower dimensionality and hence the effective band width or the kinetic energy scale is smaller than those in the bulk layers. As a result, the effective interaction scale compared to the kinetic energy is bigger at the boundary, which leads to the earlier onset of the magnetic order when the interaction $U$ increases and a generally stronger magnetic order near the boundaries.

\begin{figure}
\centering
\includegraphics[trim= 0mm 40mm 0mm 40mm, clip, width=\linewidth]{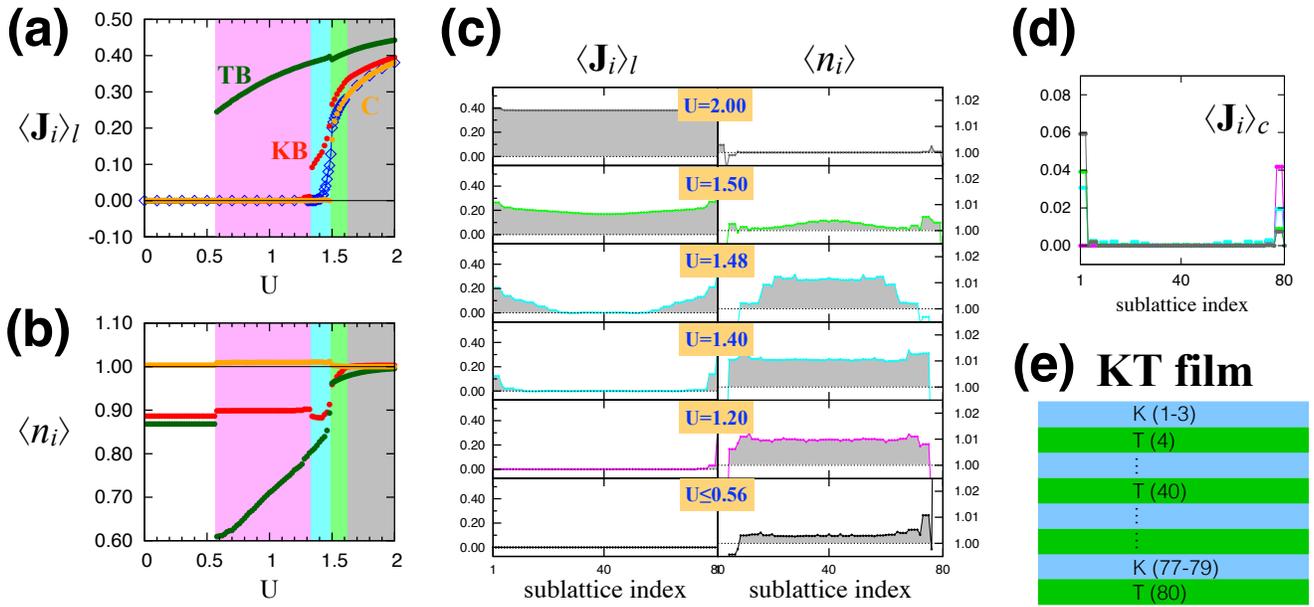}
\caption{Magnetic structure of the KT film.
(a) The local axis component $\langle {\bf J}_i \rangle_l$ of the magnetic order parameter. 
The plot shows $\langle {\bf J}_i \rangle_l$ as a function of $U$ at the triangular boundary (TB; green), kagome boundary (KB; red) and central layer (C; orange) in the film.
(b) The electron density $\langle n_i \rangle$ with the same color scheme in (a).
(c) Spatial profiles of $\langle {\bf J}_i \rangle_l$ and $\langle n_i \rangle$ over the layers (sublattices) in the KT film.
(d) Spatial profiles of the canting component $\langle {\bf J}_i \rangle_c$ of the magnetic order parameter for $U$=1.20 (pink), 1.48 (cyan), 1.50 (green), 2.00 (gray).
(e) Sublattice index for the KT film with twenty bilayers.
\label{fig:ktfilm1}}
\end{figure}

This trend is corroborated by the inhomogeneous distribution of the electron density $\langle n_i \rangle$ across the layers as shown in Figs. \ref{fig:kkfilm1}, \ref{fig:ttfilm1}, and \ref{fig:ktfilm1}.
In the paramagnetic SM state or in the weak correlation regime, the electron density is higher (lower) near the central (boundary) layers as the electrons try to maximize their 
kinetic energy gain [orange and red in Fig. \ref{fig:kkfilm1} (b)].
As the interaction $U$ increases, the energy penalty of double occupancy makes the electron density to be more and more evenly distributed
and it finally reaches $\langle n_i \rangle=1$ everywhere in the strong correlation regime.
In the intermediate correlation regime where the MM states occur, the electrons are still concentrated near the non-magnetic central layers compared to
the magnetically-ordered boundary layers, as shown for the cases of $U=1.34,~1.48,~1.50$ in Fig. \ref{fig:kkfilm1} (c).
Notice that this result is the consequence of the fully self-consistent computations of the magnetic order parameters and electron density, which allow the layer-by-layer variation. 

Similar patterns are also found in the TT and KT film structures as summarized in Figs. \ref{fig:ttfilm1} and \ref{fig:ktfilm1}.
In the TT film, the onset of the magnetic order at the boundary layers occurs at a smaller value of $U$ compared to the KK film (compare Fig. \ref{fig:ttfilm1} with \ref{fig:kkfilm1}),
which is due to the even less number of hopping paths or an even smaller band width of the triangular boundary layer [see Fig. \ref{fig:filmstructure} (a)].
As the KT film has both of the kagome and triangular boundary layers, there exit two separate onsets of the magnetic order at the boundaries, namely the magnetic
order develops first at the triangular boundary and then at the kagome boundary before it spreads to the central layers (see Fig. \ref{fig:ktfilm1}).

\begin{figure}
\centering
\includegraphics[trim= 0mm 80mm 0mm 0mm, clip, width=\linewidth]{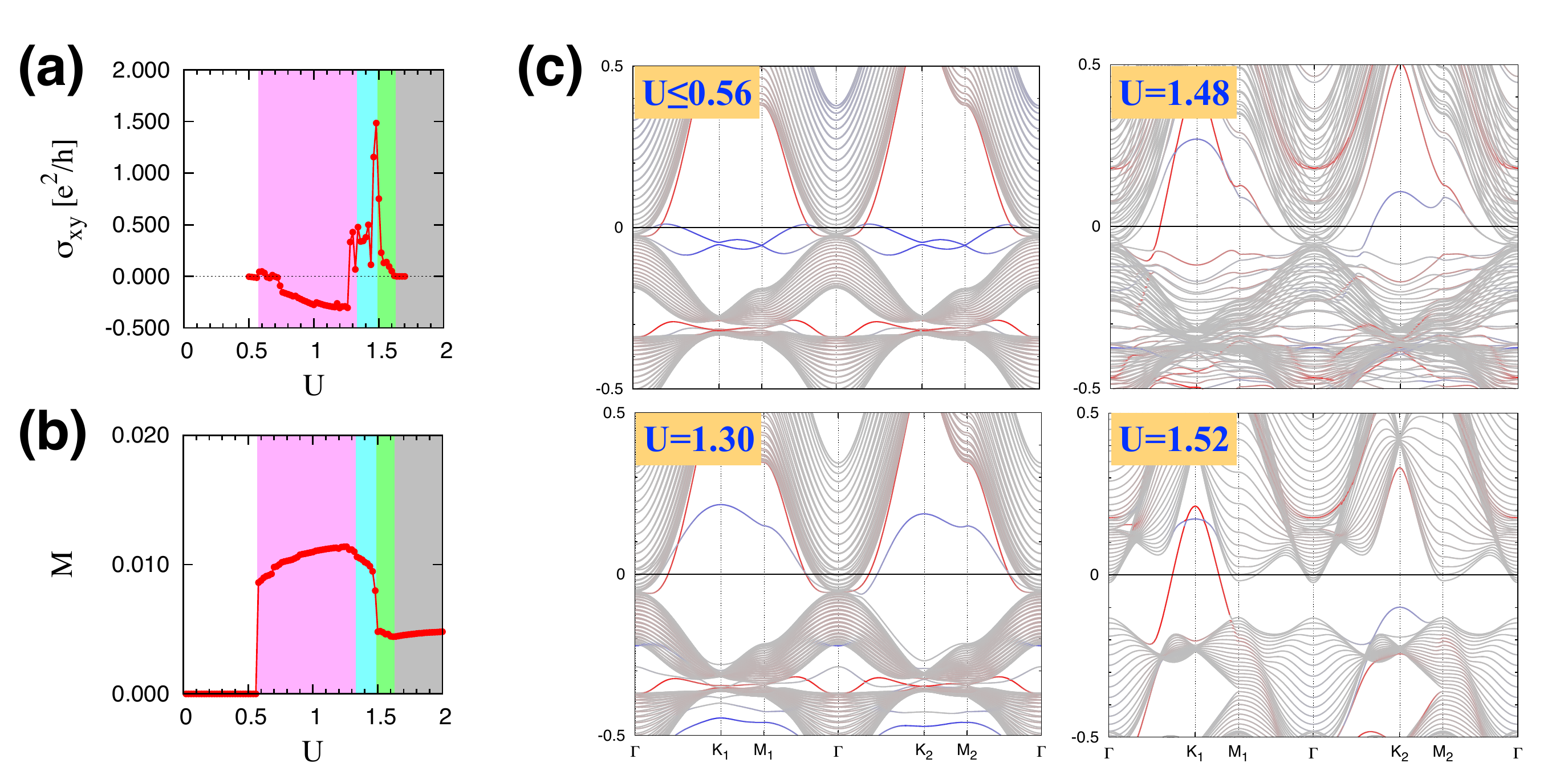}
\caption{Electronic structure of the KT film.
(a) The Hall conductivity $\sigma_{xy}$ as a function of $U$.
(b) The magnetization $M$ along the [111] direction as a function of $U$. 
(c) Electron band structures of the SM ($U\leq0.56$), MM-TB ($U=1.30$), MM-KB ($U=1.48$), and MM-C ($U=1.52$) phases.
In each plot, the surface modes localized at the top kagome and bottom triangular layers (red and blue, respectively) are distinguished from the bulk bands (gray).
\label{fig:ktfilm2}}
\end{figure}

In the TT film, only one of the MM phases, MM-TB (marked with pink shade in Figs. \ref{fig:ttfilm1} and \ref{fig:ttfilm2}), occurs.
This is due to the fact that the surface bands from the triangular layer are very flat [see color lines in Fig. \ref{fig:ttfilm2} (c)] in contrast to the case of
the kagome boundary in the KK film, where the dispersive surface bands allow the appearance of the MM-C phase with a negative indirect gap.
In the KT film with both types of the boundary layers, all the three magnetic metal phases, MM-TB (pink), MM-KB (cyan), and MM-C (green) with 
the corresponding behaviors of the Hall conductivity arise (Figs. \ref{fig:ktfilm1} and \ref{fig:ktfilm2}).
In addition, the electronic structures have two distinct surface bands from the kagome and triangular boundaries [see red and blue curves in Fig. \ref{fig:ktfilm2} (c)].

As discussed in the previous section, the large anomalous Hall effect in the MM phases arises from a large of number of crossings of non-degenerate bands, which lead to the finite Berry flux via the associated monopole structures in the momentum space, and the broken cubic symmetry of the film. 
Hence one can regard these phases as descendants of the Weyl semimetal phase in the bulk system. 

Now we discuss the crucial difference between the results of our work and the previous theoretical studies on thin film structures of the pyrochlore iridates.
In Ref. \citeonline{yang2014}, similar thin film structures were considered, but it was assumed that the magnetic order and electron density are uniform across the layers. 
This leads to qualitatively different behaviors of the magnetic and electronic structures. We show that the fully relaxed self-consistent 
computations give rise to the non-uniform magnetic order and electron density across the layers. 
In the previous work\cite{yang2014}, the presence of the so-called ``hidden topological phase" with the band crossing of gapless surface states 
was attributed as the origin of the largest anomalous Hall effect.
Such a phase would occur only in the KT film. 
In our work, it is shown that the largest anomalous Hall effect can occur for all of the boundary conditions and 
it comes mainly from the large number of crossings of the bulk non-degenerate bands and the associated large Berry flux,
which result from the inhomogeneous effective fields of the non-uniform magnetic order across the layers.
These predictions may be tested by studying the pyrochlore thin flims with different choices of the boundary/surface layers.

In Ref. \citeonline{hu2015}, the authors considered a ultra-thin limit, namely one unit of the double layer made of one kagome and one triangular lattices
and the triple-layer made of two kagome and one triangular lattices. The magnetic order in these cases is very different from the AIAO order of the bulk system and the overall phase diagram is quite different from the thin film geometries (20 bilayer unit) considered in our work. Hence the physics in this ultra-thin limit has no connection to the bulk properties of the pyrochlore iridates. Clearly, the thin film systems that we are considering here are more naturally connected to the bulk system.

Very recently, the fabrication of the Eu$_2$Ir$_2$O$_7$ thin film has been reported.\cite{fujita2015,fujita2016,fujita2016_2nd} We expect future experiments on these systems would shed
significant light on the nature of emergent quantum phases in the pyrochlore iridates.

\section*{Acknowledgements}

We are grateful to Bohm-Jung Yang and Kun Woo Kim for helpful discussions. 
This work was supported by the NSERC of Canada, the Canadian Institute for Advanced Research, and the Center for Quantum Materials
at the University of Toronto. This research was also supported in part by Perimeter Institute for Theoretical Physics. 
Research at Perimeter Institute is supported by the Government of Canada through Industry Canada and 
by the Province of Ontario through the ministry of Research and Innovation.
Computations were performed on the gpc supercomputer at the SciNet HPC Consortium. SciNet is funded by: the Canada Foundation for Innovation under the auspices of Compute Canada; the Government of Ontario; Ontario Research Fund - Research Excellence; and the University of Toronto.

\section*{Author Contributions}

Y. B. K. conceived of the project, and K. H. and Y. B. K. performed the calculations and wrote the manuscript.

\section*{Additional Information}
\textbf{Competing financial interests:} The authors declare no competing financial interests.

\end{document}